\documentclass[aps,pre,preprint,superscriptaddress]{revtex4}
\usepackage{amsmath,amssymb}
\usepackage{graphicx}
\usepackage{color}

\newcommand{\xz}[1]{\textcolor{black}{ #1}}

\begin{document}

\title{Free volume under shear} \author{Moumita
  Maiti}\affiliation{Institute for Theoretical Physics, Georg-August
  University of G\"ottingen, Friedrich-Hund Platz 1, 37077
  G\"ottingen, Germany} \author{H. A. Vinutha}\affiliation{Jawaharlal
  Nehru Center for Advanced Scientific Research, Jakkur Campus,
  Bengaluru -- 560064, India} \affiliation{TIFR Center
  for Interdisciplinary Sciences, 21 Brundavan Colony, Narsingi,
  500075 Hyderabad, India}\author{Srikanth Sastry}
\affiliation{Jawaharlal Nehru Center for Advanced Scientific Research,
  Jakkur Campus, Bengaluru -- 560064, India} \affiliation{TIFR Center
  for Interdisciplinary Sciences, 21 Brundavan Colony, Narsingi,
  500075 Hyderabad, India}\author{Claus Heussinger}
\affiliation{Institute for Theoretical Physics, Georg-August
  University of G\"ottingen, Friedrich-Hund Platz 1, 37077
  G\"ottingen, Germany}
\begin{abstract}

  Using an athermal quasistatic simulation protocol, we study the
  distribution of free volumes in sheared hard-particle packings close
  to, but below, the random-close packing threshold. We show that
  under shear, and independent of volume fraction, the free volumes
  develop features similar to close-packed systems -- particles
  self-organize in a manner as to mimick the isotropically jammed
  state. We compare athermally sheared packings with thermalized
  packings and show that thermalization leads to an erasure of these
  structural features.
  The temporal evolution, in particular the opening-up and the closing
  of free-volume patches is associated with the single-particle
  dynamics, showing a crossover from ballistic to diffusive behavior.

  \end{abstract}


%

\date{\today}

\maketitle

\section{Introduction}
The transition from a flowing liquid to a jammed solid state has been
a subject of research in a wide range of systems, including granular
matter, colloidal suspensions, and diverse types of glass formers, and
in the context of gelation. A particular context, in which the
geometry associated with such a transition is important, has been that
of the jamming to unjamming transition in athermal particle packings,
often modeled using sphere packings, with or without the presence of
external driving. The relationship between the jamming phenomenology
and thermal systems, either the rheology of driven thermal systems, or
to the glass transition in undriven systems, are intensely
investigated currently \cite{liu-nagel-1998,liu-nagel-2010,torquatorev}.

Disordered assemblies of spheres undergo a {\it jamming} transition at
a packing fraction of $\sim 0.64$, at which the pressure for hard
sphere packings diverges. 
At this {\it random close packing} (RCP) density, individual hard
spheres are entirely constrained by their neighbours and have no space
to move around. The network of such spheres in contact (``backbone'')
spans the system and may thus support external load.

The geometrical properties of sphere packings and their influence on
the mechanical response is a complex problem with many different
facets. One key observable is the connectivity of the network which is
isostatic at RCP \cite{moukarzel,ohern2003,wyart2005}. Wyart~\cite{PhysRevLett.109.125502}
discussed the consequences of opening of contacts, as well as the
distribution of gaps, i.e. distances between non-contacting
particles. Atkinson et al. \cite{atkinson13:_detail} have studied the
structure of rattlers, the particles that are not constrained by
enough contacts. Schr\"oder-Turk et al.~\cite{schroeder-turkEPL2010}
have observed a signature of RCP in the shape of Voronoi cells of the
particles.

Packings at RCP also have special mechanical properties.
The bulk modulus, for example, is finite for soft-sphere packings at
RCP, while it generically vanishes for spring networks of equivalent
connectivity~\cite{ellenbroek09:_non,PhysRevE.85.021801} Apparently,
the particles in a packing can \emph{organize} in order to resist
compressive forces, in a way that is not possible for spring networks.

In the vicinity of random close packing, spheres have finite free
space to move, and free space vanishes approaching { RCP}. For
thermal systems this implies slow dynamics approching { RCP}. At
the same time, dynamics in sheared systems at zero temperature speeds
up by approaching RCP. Indeed, particles in shear flow are found to
move ever faster, the less space they
have~\cite{heussingerEPL2010}. It is an interesting but essentially
unsolved question how such non-trivial dynamics arises and couples to
other properties of the system, in particular to the singular behavior
of the correlation length or the rheological
coefficients~\cite{PhysRevLett.109.108001,PhysRevLett.107.158303,otsukiPRE2009,Lerner27032012,PhysRevLett.109.105901}. Here,
we are interested in the analysis
of the free space available for particle movement, and how that may be
affected by externally imposed deformation, and in turn how such
changes in geometry influence particle flow.


\begin{figure*}[t]
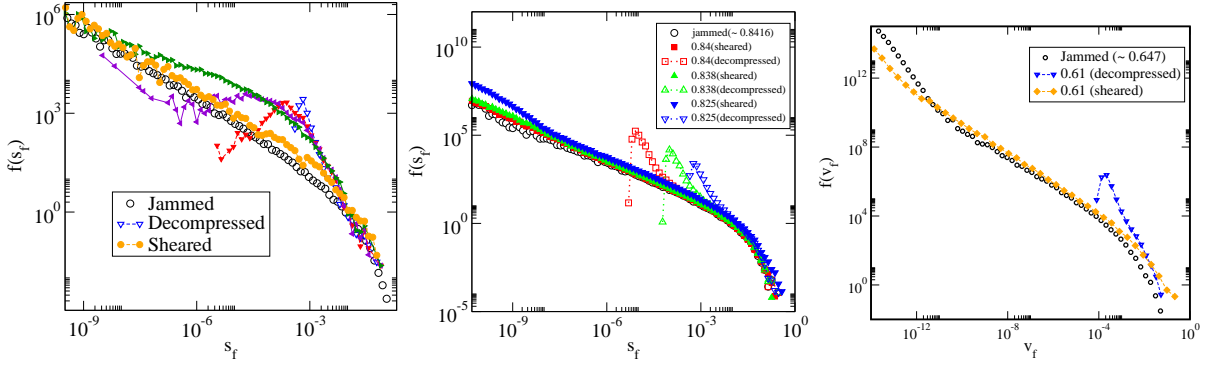
 
   \centering
   \includegraphics[width=2.05in]{fig1a.eps}
   \includegraphics[width=2.1in]{fig1b.eps}
   \includegraphics[width=2.0in,clip=true]{fig1c.eps} 
   \caption{Probability distribution of free volumes (2D): (a)
     $\phi=0.825$ for decompressed (unsheared) configurations and for
     different elapsed strains; (b) Different volume fractions in the 
     steady state and compared with the distribution for nearly jammed
     configurations. (c) Data  for the 3D system.
}
   \label{fig:distr.free}
\end{figure*} 

\xz{The divergence of pressure in a hard particle system is  a consequence of the
  vanishing of free space. Pressure is the ratio of free surface to
  free volume~\cite{speedy}, the latter vanishes at jamming.
In the fluid phase, the free volume distribution is peaked around the
mean free volume, and at jamming the distribution can be expected to
have a delta function peak at zero free volume for the backbone
spheres.
  Indeed close to jamming, for isotropic, thermalized, ensembles of
  configurations at any given density, the free volume
  distribution~\cite{maiti2014jcp} has a strong peak around the mean
  free volume. A surprising observation, however, is that there is a
  power law tail in the free volume distribution in addition, which
  appears to be a signature of being close to the jamming point. In
  the present work, we address the question of how the free volume
  distributions of sphere packings close to the jamming point get
  modified under shear deformation, both in the athermal and thermal
  cases, and what we may deduce regarding the organization of spheres
  in response to external shear.}











\section{Simulation Details}

We study two- (2D) and three-dimensional (3D) systems of $N$ soft
frictionless spheres. To avoid crystallization, we use two different
particle sizes; $\frac{N}{2}$ spheres have diameter $d$ and
$\frac{N}{2}$ have diameter $1.4d$. The particle density is quantified
via the fraction of volume (area in 2D, although we use `volume' in
the 2D case as well, when no confusion is caused) $\phi$ that is
occupied by the particles.


Particles interact via elastic repulsive forces. Two particles repel
each other with a harmonic potential energy,
\begin{eqnarray}\label{eq:}
  {E}_{\rm el} = -\epsilon(1 - \frac{r_{ij}}{d_{ij}})^2\,, \qquad r_{ij} < d_{ij}
\end{eqnarray}
where $r_{ij}$ is the distance between the two particles. The cut-off
$d_{ij}=(d_i+d_j)/2$ is set by the diameters of the two interacting
spheres.


To implement the shear a quasistatic simulation protocol is
used~\cite{heussingerPRL2009,heussinger2010SoftMatter}. In the
quasistatic limit (small-strain rates, $\dot{\gamma} \rightarrow 0$)
the average overlap $\delta = \langle 1-\frac{r_{ij}}{d_{ij}}\rangle$,
vanishes and the particles effectively behave as hard-spheres
(equivalent to $\epsilon\to\infty$). In recent
work~\cite{PhysRevLett.109.105901,maiti2014pre} it was shown that the statistics of
particle velocities obtained from the quasi-static simulations is
identical with the small-strain rate (Newtonian fluid) limit of fully
dynamic molecular dynamics simulations. Thus, our quasistatic results
should be considered representative for the Newtonian flow-regime of
dense suspensions close to jamming.

The jamming density under shear is at $\phi_c = 0.647$ ($\phi_c=0.842$
in 2D).
We probe densities below this limit in the range $\phi=0.61-0.64$
($\phi = 0.825\ldots 0.840$ in 2d). In this range of densities
cooperative effects set in and the correlation length increases
roughly by a factor of ten~\cite{heussingerEPL2010}.



\section{Results}

Motivated by recent results on the free volume distribution in
isotropically jammed packings~\cite{maiti2014jcp}, we ask how steady
shear affects these distributions. The free volume of a particle is
the volume (area in 2D) that is available for the center of the
particle with all other particles are fixed in space.

Starting with a jammed configuration at $\phi_c$, we decompress the
configuration to the target density at $\phi<\phi_c$ by reducing the
diameters of all particles, $d_i\to\alpha d_i$, where $\alpha$ is the
scaling factor for the particle diameters. In this way, all particles
acquire a certain amount of free volume $v_f$ (or free surface $s_f$
in the two dimensional case) with a probability distribution,
$P(v_f)$, that is highly peaked at $v_f\sim \alpha^d$ (see
Fig.~\ref{fig:distr.free} (a) for the 2d system, blue line/downward
triangles). In this figure we also highlight the distribution of free
volume for the nearly jammed configuration (before decompression),
which is a power-law (black open circles)~\cite{maiti2014jcp}.

The ``decompressed'' configuration serves as starting point to our
shear simulation. Fig.~\ref{fig:distr.free} (a) indicates the
evolution of the probability distribution as a function of the elapsed
strain. \xz{After only small amounts of strain, all particles retain
  their finite free volume, but as straining goes on the number of
  those particles reduce and a peak at zero free volume starts
  developing (not shown in the figure). At the same time, a broad tail
  develops. } In the steady-state, thus, the form of the distribution
is rather different than in the decompressed configuration. Instead of
a narrow, peaked distribution we obtain \xz{a strong peak at zero free
  volume} with a broad power law tail.

Fig.~\ref{fig:distr.free} (b) illustrates that, close to jamming, the
power-law tail is independent of density $\phi$, which is also
identical in the exponent to the isotropically jammed state. \xz{The
  same phenomenon is observed also for the three-dimensional system,
  as shown in Fig.~\ref{fig:distr.free} (c), illustrating that the
  outcome is not dependent on dimensionality}.  Apparently, during
shear particles self-organize in such a way as to mimick the
isotropically jammed state -- even though the system is not jammed and
flows as a Newtonian liquid.

\begin{figure}[t] 
   \centering
   \includegraphics[width=2.5in]{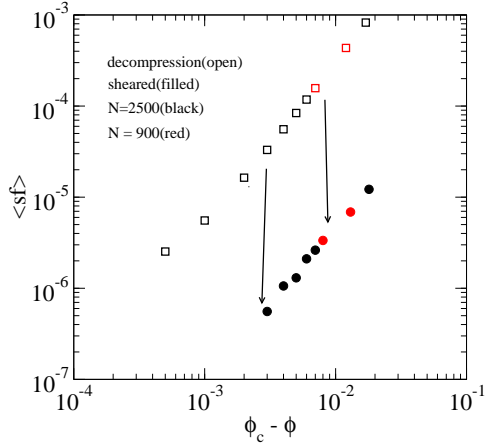}
   \caption{Average free volume as a function of volume
     fraction. Comparison of configurations under shear and
     decompressed states. Two different system sizes are compared to
     exclude finite-size effects.}
   \label{fig:avg.free}
\end{figure}

It is worthwhile comparing the situation of steady shear with that of
cyclic shear. Experiments performed at low
densities~\cite{pine2005Nature,corteNatPhys2008} have shown that after
sufficient shear cycles the system self-organizes into an
``absorbing'' state, where particles do not interact anymore. Their
trajectories during a shear cycle are strictly reversible. This
indicates that particles generate free volume around themselves in
such a way that during their shear-induced oscillations no
interactions with other particles occur. Under steady-shear the
organization apparently is opposite, and free volume is not generated
but destroyed (Fig.~\ref{fig:avg.free}). The average free volume under
shear is roughly two orders of magnitude smaller compared to the
decompressed configuration at the same density.

\begin{figure}[t] 
   \centering
   \includegraphics[width=2.5in]{fig3a.eps}
   \includegraphics[width=2.5in]{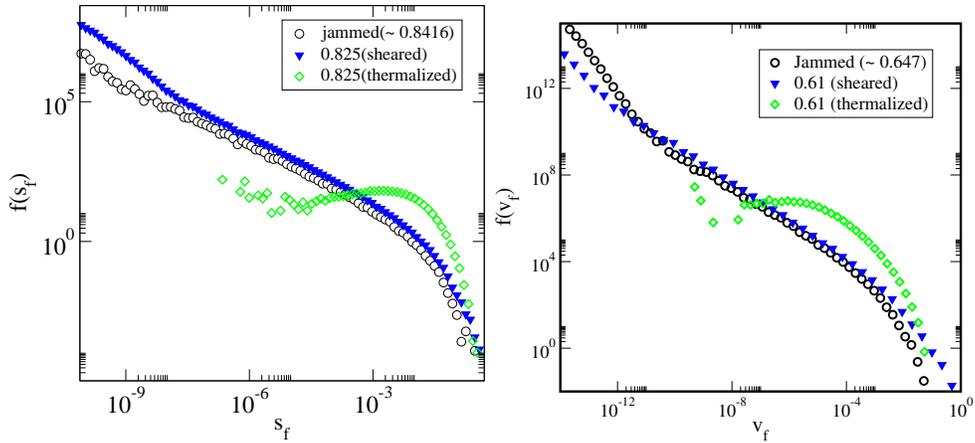}
   \caption{Comparison of free-volume distributions from jammed,
sheared 
and thermalized
configurations (top: 2D, bottom: 3D).
}
   \label{fig:free.compare.thermal}
\end{figure}

Another illustrative comparison is with thermal systems. Recall that
the system we study is inherently athermal. Particles only move
because they are driven by shear. To generate a thermalized system at
the same density we again use the decompressed state as starting
configuration, but now run short Metropolis Monte-Carlo
simulations. The resulting free volume distribution is presented in
Fig.~\ref{fig:free.compare.thermal} and compared to the sheared state.

It is clear that thermalization acts rather differently than shearing
and does not produce a power-law tail in the free volume distribution
(in agreement with Ref.~\cite{maiti2014jcp}). This is interesting also
from the point of view of shear-induced effective
temperatures~\cite{onopre2002,haxtonprl2007,bouchbinderpre2007,berthierjcp2002},
which rely on the assumption that driving by shear is in some sense
equivalent to thermal driving but at an ``effective'' temperature. At
least for the observable under consideration -- the free-volume
distribution -- this equivalence does not seem to hold here.

Another difference between the sheared and thermal systems is the
distribution of the total free volume among the particles. In the
thermal system (and also under cyclic shear) the free volume is
distributed basically among all particles. In the sheared system, only
a certain fraction of particles have a non-zero free volume. In the
density range we studied, we find a fraction of $6\%$ to $11\%$ of
such finite free volume particles. The majority of particles belong to
a backbone of locally jammed particles with zero free volume. The full
free-volume distribution therefore consists of a power-law tail for
particles with finite free volume plus a delta-function peak for the
backbone particles. As the density is increased, more and more finite
free volume particles are incorporated into the backbone, such that at
$\phi_c$ the backbone becomes globally jammed and the system
solidifies.

\begin{figure}[t] 
   \centering
   \includegraphics[width=2.5in]{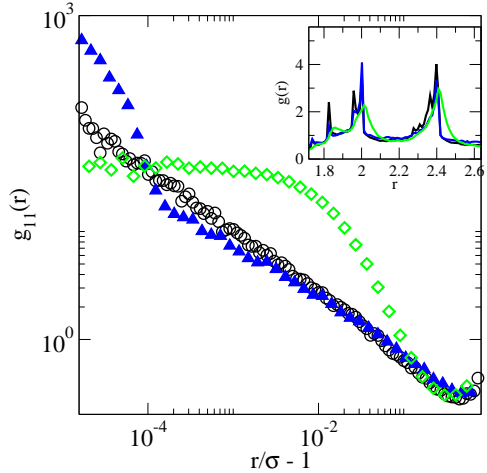}
   \caption{Pair correlation function $g(r)$ for jammed packing, as
     well as sheared and thermalized configurations (color code as in
     Fig.~\ref{fig:free.compare.thermal}). Sheared configurations
     display the same characteristic power-law slope as the jammed
     configuration. Thermalization destroys this power-law. Inset
     shows that sheared configurations have a split-second peak as
     seen in jammed configurations, whereas for thermalized
     configurations, the sharp features are smeared out.}
   \label{fig:gofr}
\end{figure}

A structural self-organization via shear flow is also visible in the
pair-correlation function. \xz{The power law singularity as well as
  the splitting of the second peak are commonly taken as the signature
  of the jammed state. Fig.~\ref{fig:gofr} shows that sheared
  configurations exhibit a similar power-law singularity with the same
  exponent as in the jammed state, whereas thermalization destroys the
  divergence. Also the splitting of the second peak in steady-shear is
  as sharp as in the jammed state but it is smeared out under
  thermalization. We have also computed the radial distribution
  function only for the finite free volume particles and find it to be
  similar as for particles with zero free volume. This is consistent
  with the findings of Atkinson {\it et al.}~\cite{atkinson13:_detail}
  who report a certain amount of spatial correlations in the
  rattlers.}

Solidification is known to occur when the backbone is exactly
isostatic. We can extract the approach to isostaticity also from the
free volumes. To this end we shrink the particles (in the steady-state
shear configurations) by a very small factor $f$. This generates a
finite free volume $v={\cal O}(f^d)$ for the backbone particles. The average
number of facets of these free volume patches then defines the
connectivity of the backbone.

\begin{figure}[t] 
   \centering
   \includegraphics[width=2.5in]{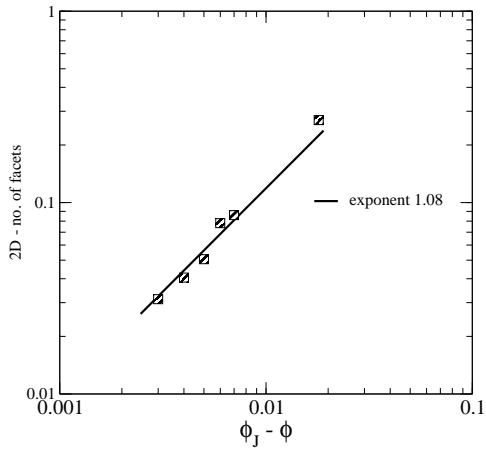}
   \caption{Effective connectivity in the backbone of zero-free volume
     particles -- measured via the number of facets of the free volume
     after tiny amount of decompression.}
   \label{fig:facets}
\end{figure} 

\xz{Fig.~\ref{fig:facets} shows that the backbone network is
  hypostatic (as expected in the fluid regime below jamming) and thus
  not mechanically stable. Isostaticity is approached as
  $z_{\rm iso}-z\sim (\phi_c - \phi)^x$, with an exponent $x$ close to
  one.  Being hypostatic, there are a number of zero frequency modes
  along which particles can move without cost in energy. This motion
  is, for example, visible as a short-time ballistic regime in the
  mean-square displacement~\cite{heussingerEPL2010}.  Similarly, this
  motion is reflected in the temporal evolution of the free volume of
  the finite free volume particles. The locations of such particles
  represent holes in the network of backbone particles and how these
  holes close or open-up reflects how particles move in the network.}

\begin{figure}[t] 
   \centering
   \includegraphics[width=2.5in]{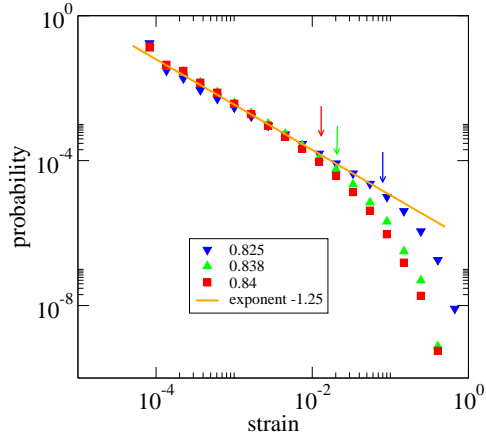}
   \caption{The distribution of strain values needed for closing a
     hole. The arrows signify the strain $\gamma_{\rm msd}$, at which
     the particle dynamics (as measured by the mean-square
     displacement, msd) crosses over from ballistic to diffusive
     (taken from Ref.\cite{heussingerEPL2010}).}
   \label{fig:time}
\end{figure} 

We calculate the strain needed to close a hole. This represents a
measure for the ``life-time'' of finite free volume for a particle
that resides in this hole. The probability distribution of strains to
close the holes is displayed in Fig.~\ref{fig:time}. The striking
feature is its power-law tail, which is a consequence of the broad
distribution of hole sizes. Holes can close after arbitrarily small
strains.
On large strain-scales, the tail is cut-off at strains $\gamma_c$ that
decrease with increasing the volume-fraction towards $\phi_c$.
In fact, the cut-off $\gamma_c$ is comparable in magnitude to the
strain $\gamma_{\rm msd}$, at which the mean-squared displacement has
a crossover from ballistic to diffusive behavior (extracted from
Fig. 1 in Ref.~\cite{heussingerEPL2010}). The strains
$\gamma_{\rm msd}$ for three different $\phi$ are indicated in
Fig.~\ref{fig:time} by arrows. This connects the evolution of
free-volume patches -- holes in the network of backbone particles --
with the single-particle dynamics. The cross-over to diffusion signals
the onset of the decorrelation of the floppy, zero-energy modes, and
this decorrelation is associated by the opening and closing of these
holes.



\section{Conclusion}

In conclusion, we have shown that \emph{unjammed athermal hard-particle 
  systems under shear} self-organize such that structural properties
resemble those of \emph{jammed packings}. In particular, we have
reported the probability distribution of free-volumes as well as the
pair-correlation function.

The free-volume distribution in (nearly-)jammed packings has
previously been shown to display a power-law
tail~\cite{maiti2014jcp}. We show here that this tail is also present
in sheared systems in a finite interval of densities below the jamming
threshold. By way of contrast, thermalized configurations in this
density range generically show a peaked and rather narrow distribution
of free volumes.

Apparently, shearing and thermalization act rather differently what
regards these local structural properties. Shearing drives the system
``towards'' the jamming point, while thermalization drives the system
away from it. 

The average free volume in the sheared configurations is roughly two
orders of magnitude smaller than in the thermalized
configurations. Moreover free volume is heterogeneously distributed
among only a few finite free volume particles, with the majority of particles
belonging to a locally jammed backbone with zero free volume.

Our results raise the question about the relevance of finite free volume 
particles for the flow properties close to jamming. Indeed, a
power-law in the free-volume distribution means that there are
particles with arbitrarily-small amounts of extra space. As a
consequence, they can easily be integrated into the backbone after
only infinitesimal amounts of strain.

\begin{acknowledgments}
  We acknowledge financial support by the German Science Foundation
  via the Emmy Noether program (He 6322/1-1).
\end{acknowledgments}


\end{document}